\newcommand\dns{D$_{\rm n} - \sigma$}
\newcommand\kms{$\rm\,km\,s^{-1}\,$}
\newcommand\kmsm{km s$^{-1}$ Mpc$^{-1}$}
\newcommand\etal{{\it et al.}~}

\documentstyle[11pt,aaspp4]{article}
\singlespace

\lefthead{Scodeggio, Giovanelli \& Haynes}
\righthead{Distance to A2634}

\begin{document}

\title{
The  Relative Distance between the Clusters of Galaxies \\ A2634 and Coma
\footnote{
Work based in part on observations obtained at the Arecibo (AO), Kitt Peak 
(KPNO) and Palomar (PO) Observatories. Observations at the PO were made as 
part of a continuing collaborative agreement between the California Institute 
of Technology and Cornell University. The AO is part
of the National Astronomy and Ionosphere Center, which is operated by 
Cornell University under a cooperative agreement with the National Science 
Foundation. The KPNO is part of the National Optical Astronomical 
Observatories, which are operated by Associated Universities, Inc. under a 
cooperative agreement with the National Science Foundation.}
}

\author{Marco Scodeggio, Riccardo Giovanelli and Martha P. Haynes}
\affil{Center for Radiophysics and Space Research and National Astronomy
and Ionosphere Center, Cornell University, Ithaca, NY 14853\\
scodeggi, riccardo, haynes@astrosun.tn.cornell.edu}

\begin{abstract}
The Tully-Fisher (TF) and Fundamental Plane (FP) relations are used to obtain 
two independent estimates of the relative distance between the clusters A2634 
and Coma. Previously published studies of A2634 showed a large discrepancy 
between the distance estimates obtained with the TF and the \dns\ relations,
questioning the reliability of redshift-independent distances
obtained using these relations. Because of the importance of this issue, we
have obtained new distance estimates for A2634, based on much larger samples 
than previously used, and selected according to rigorous membership criteria. 
New I band CCD photometry for 175 galaxies, new 21 cm observations of 11 
galaxies, and new velocity dispersion measurements for 62 galaxies are used 
together with previously published data in building these samples.

As part of a larger project to compare the TF and FP distance-scales, we have 
obtained a new FP template using for the first time I band photometry. 
The template is derived using a sample of 109 E and S0 galaxies that
are members of the Coma cluster. Its parameters are in very good agreement with
recent determinations of the FP obtained at shorter wavelengths.
The uncertainty with which the FP can provide peculiar velocity 
estimates for single galaxies is $\simeq$0.43 mag in the distance
modulus, or 20\% of the distance. This uncertainty is slightly larger than 
the typical uncertainty that characterizes TF estimates. However this 
disadvantage is partly compensated by the fact that the sample incompleteness 
bias has a less severe effect on FP cluster distance estimates than it has on 
the corresponding TF distance estimates. Also, cluster membership is more 
readily established for early--type objects than for spirals.

After the appropriate corrections for sample incompleteness have been taken
into account, we find the TF and FP distance estimates to be in good 
agreement, both indicating that A2634 has a negligibly small peculiar 
velocity with respect to the Cosmic Microwave Background reference frame. 
Because of the high accuracy with which the two distance estimates have been 
obtained, their agreement strongly supports the universality of the 
TF and FP relations, and therefore their reliability for the estimate of
redshift-independent distances.

\end{abstract}

\section{Introduction}

The measurement of redshift-independent distances, together with that of 
redshift measurements, constitutes one of the basic tools of observational 
cosmology.  While redshift surveys alone can be used to obtain maps
of the distribution of luminous matter in the local Universe (see Giovanelli \&
Haynes 1991 for a recent review), only redshift-independent distances can
dynamically constrain the total matter content (i.e. dark plus luminous) of 
the Universe. 
In the absence of perfect ``standard candles'', i.e. objects of well
determined and invariant intrinsic properties, many different methods have been
developed to climb the ``distance scale ladder'' out to cosmologically 
significant distances (see for example Rowan-Robinson 1985). The ideal method 
would provide high accuracy, i.e. it would exhibit small dispersion in the 
predicted distance, it would be applicable over a large range of distances and 
for a well distributed set of targets, and it 
would have a well understood physical base. So far no technique has been 
developed that satisfies all of the above requirements. 
Among the available methods (see Jacoby \etal 1992 for a review), those  
most extensively utilized are the Tully-Fisher (TF) relation (Tully \& Fisher 
1977), which is applied to spiral galaxies, and the \dns\ or Fundamental Plane 
(FP) relation (Dressler \etal 1987; Djorgovski \& Davis 1987), which is
applied to elliptical and lenticular galaxies.

TF, FP and \dns\ are very similar methods, all based on the correlation 
between a photometric distance-dependent parameter and a spectroscopic, 
distance-independent quantity. The three relations have fair accuracy 
(the uncertainty in the distance prediction of a single galaxy is 
12 -- 25\%), and they can be applied over a 
large distance interval (from the Local Group out to galaxies with 
redshift greater than 0.1; see for example Franx 1993; Vogt \etal 1993, Vogt
\etal 1996, van Dokkum \& Franx 1996). Although they can be generically 
understood in terms of virial 
equilibrium, the astrophysical details of these relations are obfuscated by 
the variance in the formation and evolution histories of galaxies. 
The calibration of the relations is thus a purely empirical exercise. 
Also, because of the uncertainty on the value of the Hubble constant, and 
because the number of nearby galaxies that could be used for an absolute 
calibration of the template relations is very small, these relations are most 
commonly used to derive only relative distances. 
In the following, the term distance is therefore used as an abbreviation for 
the more appropriate ``recession velocity corrected for peculiar motions''.

It is well known that in the absence of a perfect standard candle, distance 
measurements are subject to bias, arising from a variety of sources (see 
Teerikorpi 1984,1990,1993; Sandage 1994a,b; Willick 1994; Giovanelli \etal
1996b; and references therein). Bias corrections thus play an important role 
in the measurement of peculiar velocities. It is also possible that the 
environment might have an effect. 
For example, the template TF, \dns, or FP relation might depend on the local 
density of galaxies, and thus differ among rich and poor clusters, and 
between clusters and the field. Yet again, the intrinsic scatter about the mean
relation might be coupled with the space distribution of galaxies, complicating
the bias correction procedures.

It is then legitimate to ask whether the TF, \dns\ and FP are universal 
relations. Some attempts have been made toward quantifying this issue 
(Aaronson \etal 1986; Djorgovski, De Carvalho \& Han 1988; Guzm{\'a}n \etal 
1992; J{\o}rgensen, Franx \& Kj{\ae}rgaard 1996), producing however 
conclusions of marginal clarity.

An invalid or biased template relation will yield spurious peculiar velocities.
It has even been argued, for example, that the very evidence suggesting the 
existence of a Great Attractor (Dressler \etal 1987, Lynden-Bell \etal 
1988) might result from bias (Silk 1989).
A comparison by Mould \etal (1990) between cluster peculiar velocities
obtained by respectively using the TF and \dns\ techniques shows disagreement
which exceeds that expected from the reported uncertainties associated with the
two methods. The case of A2634, pointed out by Lucey \etal (1991a; 
hereafter L91a), is even more notorious: Aaronson \etal (1986, hereafter A86) 
applied the TF relation to find the cluster roughly at rest in the Cosmic 
Microwave Background (CMB) reference frame, while L91a used the \dns\ relation 
to obtain a cluster peculiar velocity of --3400 \kms. The difference between 
the two estimates is significantly larger than the combined uncertainty of the 
two separate results (approximately a 5$\sigma$ discrepancy), leading L91a 
to conclude that environmental effects in this cD--dominated cluster 
might be responsible for systematic bias in the properties of the early-type 
galaxy population. In such case, {\it \dns\ and FP would not be universal 
relations, and they could not be used to reliably map the peculiar 
velocity field}.

Other possible explanations for the discrepancy in A2634 should be
considered. The A86 and L91a samples have very different spatial
distributions, prompting the doubt that early- and late-type galaxies may not
sample the same dynamical unit. The L91a sample is composed of 18 E and S0 
galaxies in the core of A2634 (all galaxies are at a radial distance of 
$\leq 0.2$ deg from the cluster center), while the A86 sample is composed of 
11 spiral galaxies scattered over a large area,
several degrees to the side, that includes not only A2634 
but also the cluster A2666. Scodeggio \etal (1995, hereafter S95) 
have shown that only 7 galaxies in the A86 sample are proper members of A2634, 
while the remaining 4 are foreground objects. In addition, both the A86 and 
L91a studies failed to apply bias corrections. 

Because of the important concerns raised by the previously mentioned
results, it appeared desirable to carry out a new detailed and multi--pronged
study of A2634, which would satisfactorily answer the issues of sample
adequacy and template relation universality.
In a previous paper (S95) we presented the results of the first part of 
our study of A2634, which aimed at easing sample selection concerns.
There, an extensive redshift survey provided a detailed kinematical and 
dynamical analysis of the A2634 region. The ensuing ability to characterize 
cluster membership accurately (a) suggested that the loose sample selection 
criteria of A86 did not seriously affect their inferred peculiar velocity, and 
(b) was used to select two samples, one of E + S0 and one of spiral 
galaxies, that are used here to derive new peculiar velocity estimates of 
A2634. For the spiral sample, the motion of A2634 is referred to the template 
relation recently obtained by Giovanelli \etal (1996b; hereafter G96b), based
on 24 clusters. Since such a template is not yet available
for early-type galaxies, we refer the motion of A2634 inferred from the
FP relation to a template obtained using the Coma cluster.
For that purpose, in both A2634 and Coma we have obtained new photometric
observations, and new velocity dispersions to supplement the data already 
available in the literature. In a forthcoming work (Scodeggio, Giovanelli \&
Haynes 1996) we will present a more accurate derivation of the FP template,
based on a multi-cluster sample, and extend the comparison of the TF and FP
distance-scales to a larger sample of nearby clusters.

The remainder of this paper is organized as follows. In \S~2 we discuss 
cluster membership. In \S~3 we present our new spectroscopic and photometric 
observations. In \S~4 we briefly discuss the incompleteness bias correction. 
In \S~5 we present the new TF and FP distance estimates and in \S~6 we 
summarize the main results of this work.

Throughout the paper we parameterize distance--dependent quantities via
H$_\circ=100h$ \kmsm.

\section{Cluster membership}

Reliable assignment of cluster membership is an important criterion in the 
definition of a TF or FP sample for a cluster of galaxies. 
Because of the sparse character of the spiral population in clusters,
TF samples often fail strict membership criteria. Poor knowledge
of the cluster structure or the need to restrict to the brightest objects, has
led observers to draw targets from wide regions around cluster centers in order
to maximize target counts. In this study we aim to overcome such limitations.

The properties of Coma and A2634 are well known and we summarize them
in Table 1. Thanks to the availability of a large body of galaxy redshifts,
detailed kinematical and dynamical studies of both clusters are available.
The Coma cluster was first studied in detail by Kent \& Gunn (1982), and
redshift space caustic curves that mark the separation between the 
gravitationally bound cluster members and the unbound ``field'' have been
published, most recently by Reg{\"o}s \& Geller (1989) and Gavazzi, Randone 
\& Branchini (1995). The cluster is well separated in redshift space from both 
the foreground and the background field, and membership can be reliably 
assigned to objects to an angular separation of approximately 5 degrees 
from the cluster center. Following the criterion used by Giovanelli \etal
(1996a; hereafter G96a), we define as cluster members all galaxies within 4 
degrees from the cluster center that fall within the radial velocity limits 
defined by the Reg{\"o}s \& Geller (1989) caustic curves estimated for 
$\Omega_\circ = 0.3$. The substructure within the Coma cluster recently
discussed by Colless \& Dunn (1996) appears to be insignificant for
our discussion. The difference in distance between the merging units, as
inferred by these authors, is a negligible fraction of the distance to the
Coma cluster, and would produce, at most, only a small increase in the scatter of
of the TF and FP diagrams for this cluster.

The first in--depth redshift survey of A2634 was carried out by Pinkney \etal 
(1993). A more detailed study of the cluster structure was carried out
by S95, which also estimated redshift space caustic lines.
A2634 is located in a region of higher complexity than that containing Coma. 
The cluster A2666 (cz$_{hel}$ = 8,134 \kms) lies $3^\circ$ to the East of A2634 
(cz$_{hel}$ = 9,240 \kms); two more groups, one in the foreground 
(cz$_{hel}$ = 7,546 \kms) and one in the background (cz$_{hel}$ = 11,619 \kms), 
were identified by S95, projected respectively 1.1 and 0.6 degrees to the NE 
and to the SE of A2634. In addition, two background clusters, A2622 at 
cz$_{hel}$ = 18,345 \kms, and an anonymous cluster at cz$_{hel}$ = 37,093 \kms,
are projected to within less than 1 degree from the center of A2634. 
We define as A2634 members all galaxies within 2.5 degrees from the cluster 
center that fall within the radial velocity limits defined by the S95 caustic 
curves corresponding to $\Omega_\circ = 0.3$.

In addition to the strict cluster members, we include in our samples also a
number of galaxies that we classify as ``peripheral members''. These are 
galaxies with radial velocity very close to the cluster systemic 
velocity, but sufficiently removed from the cluster center that they do not 
satisfy a strict membership criterion. 
In the following paragraph, the term cluster members will be used to 
indicate both the strict cluster members and the peripheral ones.

\section{Observations and Data Sets}

The selection of the Coma TF sample is described in G96a.
The sample includes 41 galaxies, of morphological type between Sb and Sd, 
and with inclination greater than 45\arcdeg, that satisfy the membership 
criteria discussed in the previous paragraph.
The A2634 TF and FP samples are based on the catalog of the redshift survey by 
S95. The TF sample was selected following the same criteria as in the Coma 
case, except that it includes two galaxies with inclination lower than 
45\arcdeg, and one Sa galaxy. Two very late spiral or irregular galaxies with 
serendipitously measured widths are also included in this sample, for a total 
of 27 galaxies.
Both the Coma and the A2634 FP samples were selected to satisfy membership 
criteria. The Coma sample is composed of 109 galaxies, 71 ellipticals and
38 lenticulars, while the A2634 one is composed of 55 galaxies, 22 ellipticals
and 33 lenticulars. Three galaxies were found to show some indication of spiral
structure, and have been tentatively classified as Sa in Tables 3 and 4.

\subsection{Arecibo Observations}

In addition to the data presented in G96a, measurements for 11 spiral 
galaxies in the A2634 TF sample are included here. Velocity widths were 
obtained from 21 cm observations carried out between 1990 and 1994 with the
Arecibo 305m radio telescope. The observational setup was as described in 
Giovanelli \& Haynes (1989).
Since this sample includes galaxies significantly fainter than those
customarily observed in $21\rm\,cm$ emission, integration times averaged 
0.7 hours per object on source. The typical rms noise per channel 
ranged from 0.3 to $0.9\rm\,mJy$. All observations were taken with a 
spectral resolution of approximately 8 \kms, later reduced by smoothing by 
an amount dependent on the signal-to-noise ratio. 
The velocity width was measured using an algorithm that fits the 
rising sides of the galaxian profile with low order polynomials,
and obtains the width at a 50\% level of each of the profile horn peaks 
or single peak, depending on the line shape. We refer the reader to the 
discussion in G96a for further details.

\subsection{Kitt Peak observations}

Photometric parameters for 175 galaxies were derived from 
I band CCD images obtained with the 0.9m telescope of the Kitt Peak National
Observatory (KPNO), during 3 observing runs between April 1994 and September 
1995, plus some service observing by KPNO personnel in August 1993 (A2634 
only). The telescope was used with the f/7.5 secondary, field
corrector and T2KA CCD chip (2048 x 2048 pixels), to obtain a field of view 
of 23\arcmin\ x 23\arcmin, with a spatial scale of 0.68\arcsec\ per pixel. 
All frames were obtained with 600 seconds integration time. The median seeing 
(measured as the FWHM of the stellar light profile) for the observations 
presented here was 1.5\arcsec. The cores of the clusters were mapped with a 
mosaic of frames in a regularly spaced grid, while specific pointings were 
used for target galaxies at the clusters periphery. Observations of Landolt
fields (Landolt 1992), both at I and at R band, were repeated many times 
during each night, at airmasses between 1.2 and 2.5, to provide the 
photometric calibration.
The mean uncertainty in the zero point calibration at I band was 0.021 
magnitudes. The 1993 service observing frames were obtained in non-photometric
conditions, and have been calibrated ``a posteriori'', using overlapping
sky regions between those frames and frames obtained in photometric conditions
during the Sept. 1994 and Sept. 1995 runs. An average of 15 stars was used to
calibrate each frame, with mean uncertainty of 0.028 magnitudes. 

The data reduction process was performed using standard IRAF\footnote{IRAF
(Image Reduction and Analysis Facility) is distributed by NOAO, which is operated 
by the Association of Universities for Research in Astronomy, Inc. (AURA), under 
cooperative agreement with the National Science Foundation.} procedures, and 
the GALPHOT surface photometry package written for IRAF/STSDAS\footnote{STSDAS
(Space Telescope Science Data Analysis System) is distributed by STScI, which is
operated by AURA, under contract to the National Aeronautics and Space 
Administration.} by W. Freudling, J. Salzer, and M.P. 
Haynes. All frames have been bias-subtracted and 
flat-fielded using a ``superflat'' obtained combining, with a median filter, 
a large number ($\geq 50$) of frames. For each galaxy, a local sky 
background was determined as the mean number of counts measured in 10-12 
regions of ``empty'' sky, and subtracted from the frame. 
The typical uncertainty in this mean background, determined from the 
differences in the sky value obtained for galaxies in the same frame, is 
approximately 0.2\%. All pixels contaminated by the light of foreground stars 
or nearby galaxies, or by cosmic rays hits, were blanked, and excluded from 
the final steps of surface photometry.
A galaxy's light distribution was fitted with elliptical isophotes using a 
modified version of the STSDAS {\it isophote} package,
maintaining as free parameters the ellipse center, ellipticity and position 
angle, and incrementing the ellipse semi-major axis by a fixed fraction of 
its value at each step of the fitting procedure. The fitted parameters  
yield a model of the galaxy, which in turn is used to compute integrated 
magnitudes as a function of semi-major axis. 

The outer part of the surface brightness profiles of spiral galaxies were
fitted with an exponential disk. This fit was used to extrapolate
the disk profile to infinity, and to define the interval of radii
where the mean disk ellipticity is to be computed. The latter was then used 
to derive the galaxy inclination. Total magnitudes were obtained adding the
flux corresponding to the extrapolated part of the disk profile to the flux
measured within the outermost fitted elliptical isophote, and have 
a median uncertainty of 0.04 magnitudes. The uncertainty on the ellipticity
is a function of inclination itself, and can vary from 1 to 20\%,
as shown in G96a.
The surface brightness profile of each early-type galaxy was fitted with a 
de Vaucouleurs $r^{1/4}$ profile, yielding an effective 
radius $r_e$ and effective surface brightness $\mu_e$ (the mean surface 
brightness within $r_e$). The fit was performed from a radius equal to twice 
the seeing radius, out to the outermost isophotes for E galaxies;  for S0 and 
S0a galaxies only the central core was fitted . Total magnitudes were 
obtained by extrapolating the $r^{1/4}$ fit to infinity (E galaxies), or by
extrapolating to infinity the exponential profile that fitted the outer parts 
of the galaxy light profile (S0 and S0a galaxies), and adding the flux 
corresponding to the extrapolated part of the profile to the one measured 
within the outermost fitted galaxy isophote. The median uncertainty on the 
determination of $r_e$ is 5\%, and 0.06 mag on that of $\mu_e$.

For galaxies in close galaxy pairs, and for brightest cluster members and the 
smaller companions embedded in their halos, an iterative subtraction procedure 
was used to obtain the individual galaxy light profiles. 
First, the brighter object was fitted, after the fainter one(s) was 
blanked. The fit was then used to build a model of the brighter galaxy, which 
was subtracted from the original image. The subtracted image was used to fit 
the fainter object, and then to build a model for it; the 
latter was then subtracted from the original image and the bright object 
fitted again. The procedure converged usually at the second iteration, 
producing companion-subtracted light profiles, though the uncertainty on 
$\mu_e$ was higher than in other cases.

\subsection{Palomar Observations}

Stellar velocity dispersion measurements for 62 early-type galaxies in our FP 
samples were obtained from moderate dispersion optical spectra.
All spectroscopic observations were obtained with the Hale 5m telescope of the
Palomar Observatory, during 4 observing runs between September 1992 and 
September 1994. The red camera of the Double Spectrograph (Oke \& Gunn 1982)
was used with a 1200 lines mm$^{-1}$ grating and a 2\arcsec\ x 128\arcsec\ slit
to obtain spectra with a dispersion of 0.86 \AA\ per pixel, and a resolution 
of 2.2 \AA\ (corresponding to a velocity resolution of 129 \kms at 5300 \AA). 
The spectral coverage was approximately from 5000 to 5600 \AA, centered on the
Mg Ib lines at $\sim$5175 \AA.
Exposure times ranged from 15 to 90 minutes, depending on the brightness of the
target galaxy, with a median value of 30 minutes.  The slit was kept to 
an East-West orientation, unless a close galaxy pair was being observed, and 
both spectra were recorded simultaneously.
HeNeAr lamp spectra were obtained before and after each galaxy observation, to
provide wavelength calibration. Late G and early K type giant stars were 
used as velocity standards and as templates for the velocity 
dispersion measurements.

All spectroscopic data were reduced using standard IRAF procedures. All frames
were bias-subtracted and flat-fielded using normalized dome flat-fields.
No correction for the slit profile was necessary. Cosmic ray hits were
removed interactively from all frames. Because approximately half of the 
observations were performed in non-photometric conditions, the spectra were 
not flux-calibrated.
The wavelength calibration was obtained in two steps. First, the HeNeAr lamp
spectra frames were used to obtain a two-dimensional dispersion solution.
Typically 12-14 spectral lines, spread uniformly over the entire spectral
range, were used in the fit. The r.m.s. deviations between true and fitted
wavelength were between 0.06 and 0.1 \AA. The galaxy spectra were then
transformed to linear dispersion using the lamp dispersion solution. Next,
the two strong night-sky lines at 5460.7 and 5577.4 \AA\ were used to check
the absolute wavelength calibration. In a small number of cases where the 
original calibration was not satisfactory, the wavelength solution was rigidly
shifted to bring the two sky lines in agreement with their expected wavelength.
Finally the global consistency of the wavelength calibration procedure was
verified using the velocity standard stars. Radial velocities were measured 
repeatedly for all stars, using in turn one star as template and all others 
as unknown. The r.m.s. dispersion in the differences between true and
measured velocities, considering spectra obtained in all 4 observing runs, is 
approximately 9 \kms, in good agreement with the measured uncertainty in the
wavelength of the two sky lines.

One-dimensional galaxy spectra, to be used for the velocity dispersion 
determinations, were extracted using a 6\arcsec\ wide window (in the
cross-dispersion dimension), centered on the peak of the galaxy continuum.
All measurements of velocity and velocity dispersion were obtained using the
cross-correlation technique of Tonry \& Davis (1979), implemented in the IRAF
task {\it fxcor}. The basic assumption of this and other similar methods 
(Sargent \etal 1977, Franx Illingworth \& Heckman 1989, Bender 1990) is that 
the spectrum of an early-type galaxy is well approximated by the spectrum of
its most luminous stars (K0 -- K1 giants), modified only by the effects of the 
stellar motions inside the galaxy. Since these motions introduce only a Doppler
shift in the stellar spectra, the galaxy spectrum is given by the
convolution of the spectrum of a K giant star with the line of sight stellar
velocity distribution (LOSVD). Therefore the LOSVD can be obtained by
deconvolving the galaxy spectrum. This task was done in the
Fourier transform domain, where convolution and cross-correlation reduce to a
product. Before the Fourier transform was computed,
all spectra and templates were continuum-subtracted, normalized, end-masked
with a cosine bell function, and logarithmically re-binned.

Because of the limited signal-to-noise ratio of the spectra, and in order to 
facilitate comparisons with other studies, we constrained the
deconvolution procedure by assuming that the LOSVD is Gaussian, and therefore 
characterized by two parameters only: a mean velocity and a velocity
dispersion $\sigma$. The accuracy of the procedure was determined using 
simulated galaxy spectra. We broadened stellar template spectra with
Gaussian profiles that simulate a large range of velocity dispersions, added
Poissonian noise to the broadened spectra, and then measured the velocity
dispersion using the original spectrum as the template.
These tests showed that {\it fxcor} produces an overestimate of the velocity 
dispersion at the 4--5\%\ level, and therefore we have corrected the raw
measurements to remove this effect. 
The final velocity dispersion values were obtained by averaging the 
determinations obtained with five different stellar templates, of spectral 
types between G9 and K2.

Given the relatively large size of the extraction window, it is possible that
galaxy rotation, if present, might contribute some significant broadening of
the spectral lines, and bias the velocity dispersion measurements.
We have obtained rough estimates of the galaxy rotation velocity, and of its
contribution to the line broadening, using the original two-dimensional 
spectra. A five points rotation curve within the 6\arcsec\ extraction aperture
was obtained for each galaxy, based on 1-dimensional spectra extracted using 
2-pixel wide apertures positioned side by side. The differences in the radial
velocity measured in each one of the four lateral spectra, and the one measured
in the central one, define the projection of the rotation curve along the 
slit, which was arbitrarily oriented in the E--W direction. The contribution 
from rotation to the line broadening was estimated using a very simple model. 
A galaxy spectrum was simulated combining 5 copies of a stellar template 
spectra, broadened with a Gaussian of fixed width, and shifted to reproduce 
the 5 velocities in the rotation curve. 
The weights used in the combination were derived from the
relative intensities in the continuum of the 5 spectra used to determine the
rotation curve. The comparison of the velocity dispersion measured in the
simulated spectrum with the one used to broaden the stellar templates provides
an estimate of the amount of broadening due to rotation.

Given the accuracy with which we can measure radial velocities in our spectra,
and the small number of points used to derive the rotation curve, we cannot
reliably measure rotation velocities smaller than $\simeq$25 \kms. Within the
A2634 and Coma samples presented here (62 galaxies), 21 galaxies do not show
detectable rotation, while the remaining 41 have a median rotation velocity
of 49 \kms, with the largest measured rotation velocity being 132 \kms. 
This rotation is responsible for an average broadening of the LOSVD of 
$\simeq$2\%, with a maximum of $\simeq$7\% for the largest rotation velocity.
The velocity dispersions listed in Table 3 and 4 are corrected for this effect.

Our spectroscopic sample partly overlaps those presented by L91a for A2634,
and by Lucey \etal (1991b, hereafter L91b), Davies \etal (1987), and Faber 
\etal (1987) for Coma. We can therefore compare the different velocity 
dispersion scales, and eventually merge all spectroscopic observations in a
single database.
For Coma, we have 6 velocity dispersion measurements in common with Davies 
\etal (1987) and Faber \etal (1987), and 6 measurements in common with L91b.
In both cases the mean difference between measurements is $\simeq$1\%, with
a dispersion of $\simeq$10\%, which is close to the expected value obtained
combining the quoted uncertainties of the single measurements.
For A2634 we have 7 velocity dispersion measurements in common with L91a. Our 
measurements are systematically smaller, by 14\% on average, although much of 
this discrepancy is due to just two galaxies (330649 and 330678, L91a objects 
1482 and 134). For these two objects our velocity dispersions are 24\% smaller 
than those published by L91a. Figure 1 shows the comparison between our
measurements and those found in the literature. Filled symbols identify galaxies
in the Coma cluster, while empty symbols identify the galaxies in A2634 that we
have in common with L91a.
Given the good agreement between the different measurements, and because the
small number of objects in common with the L91a sample prevents an accurate
determination of any possible offset in the L91a velocity dispersion scale,
in the following we use the L91a, L91b, Davies \etal (1987), and Faber \etal
(1987) velocity dispersion measurements as reported in the literature, and
combine them with the new measurements presented here.

\subsection{TF and FP samples}

Both A2634 and Coma are included in the cluster sample of G96a and G96b. 
Because no new TF observations are presented here for the Coma cluster, we 
do not discuss that sample again (all the data can be found in Table 2 in 
G96a). New TF observations have been obtained for 11 galaxies in A2634, and 
we present them in Table 2, together with the sample of G96a, giving a 
complete list of the TF sample for that cluster. 
Table 2 is organized as follows: \\
Col. 1: Galaxy name, using one or two galaxy identifiers. If the galaxy is 
listed in the UGC catalog, the UGC number is listed first; if not, our 
internal coding number is given. The second identifier is the NGC or IC number,
 if the galaxy is listed in those catalogs, or the CGCG field and ordinal 
number within that field, if the galaxy is listed in the CGCG. If the galaxy 
is neither an NGC/IC nor a CGCG object, no second name is entered. \\
Cols. 2 and 3: Right Ascension and Declination, in 1950.0 epoch. \\
Col. 4: morphological type code, in the RC3 scheme (1 for Sa, 3 for Sb, 5 for
Sc, 10 for Irr). Morphological types were derived from visual inspection of
the blue plates of the Palomar Observatory Sky Survey (POSS).\\
Col. 5: Radial velocity, in \kms, measured in the CMB reference 
frame as defined by Kogut \etal (1993). \\
Col. 6: Angular distance from the cluster center, in degrees. \\
Col. 7: Membership code: A2634 members according to S95 are identified by a 
``c''; peripheral objects are identified by a ``g''; members of S95 
foreground group at cz = 7500 \kms are identified by a ``7''. \\
Col. 8: Observed velocity width, in \kms. Velocity widths are derived either 
from either 21 cm observations or from optical rotation curves, as described 
in G96a. To that source we address the reader for details on the corrections 
applied for instrumental, reduction, and relativistic broadening, turbulent 
motion, and inclination. \\
Col. 9: Velocity width, in \kms, corrected for turbulent broadening, 
instrumental and data taking effects, cosmological stretch, and shape of the 
optical rotation curve, but not for inclination. \\
Col. 10: Velocity width, in \kms, corrected for inclination. \\
Col. 11: Inclination, in degrees. A galaxy inclination is given by 
$\cos^2 i = {{(1-e_{corr})^2-q^2}\over{1-q^2}}$
where $e_{corr}$ is the mean ellipticity of the disk, corrected for the 
smearing effects of seeing (see Giovanelli \etal 1995), and $q$ is the 
intrinsic axial ratio of spiral disks (we assume $q = 0.13$ for Sbc and Sc 
galaxies, and $q = 0.2$ for all other types). \\
Col. 12: Logarithm (base 10) of the corrected velocity width, and its
associated uncertainty. The format 2.377(13) should be read as 
2.377$\pm$0.013. \\
Col. 13: Measured apparent I band total magnitude. \\
Col. 14: Corrected I band magnitude. Total magnitudes are corrected for 
extinction within our Galaxy (adopting 
Burstein \& Heiles 1978 prescriptions, and $A_I = 0.45A_B$), for the 
cosmological k-correction term (adopting Han 1992 $k_I = 0.16z$) and for 
internal extinction. This is done following the prescriptions of G96A, with 
the correction to face-on value that depends on the galaxy inclination, and 
also on its luminosity. \\
Col. 15: Absolute magnitude. This is obtained assuming that the galaxy 
is at the distance indicated by the cluster redshift, in the CMB reference 
frame, for the objects labeled ``c'' in col. 7, and at a distance 
indicated by the galaxy redshift, for the other objects, for H$_\circ$=100 
\kmsm. The magnitude uncertainty, in hundredths of a magnitude, has the same 
notation as the velocity width uncertainty in Col. 12. \\
Col. 16: If an asterisk appears in this column, special comments on the
object are available in section 3.5. \\

Figure 2 (upper panels) shows the spatial distribution of the galaxies in the
A2634 and Coma TF samples (large filled and open symbols identify proper and
peripheral cluster members, respectively). The two dashed 
concentric circles have respectively radii of 1 and 2 Abell radii, $R_A$. 
In the lower panels of Figure 2, the radial velocity is plotted versus
the angular separation from the cluster center; the
dashed vertical lines are at 1 and 2 $R_A$, the solid horizontal lines
refer to the systemic velocity of the two clusters, and the solid curves 
mark the redshift space caustic lines, taken from S95 for A2634 and from 
Reg{\"o}s \& Geller (1989) for Coma.
In both panels the small filled symbols indicate other galaxies with measured
redshift located in the area.

The galaxies in the FP samples of A2634 and Coma are listed in Tables 3 
and 4, respectively. The spectroscopic observations of the 109 Coma galaxies
in Table 4 derive principally from a large extant body of data, namely: 
51 objects from Lucey \etal (1991b, hereafter L91b), 36 S0 galaxies from
Dressler (1987), and 37 E galaxies jointly from Davies \etal (1987) and Faber 
\etal (1987). The spectroscopy of the 55 galaxies listed in Table 3, in 
the cluster A2634, includes our own observations, complemented by 18 
objects in the L91a sample. This sample of 55 objects includes 3 possible
members of the foreground group near 7,000 \kms and one in the background group
near 11,000 \kms discussed by S95. All the photometric data in tables 3
and 4 are new and obtained by us.
The contents of Tables 3 and 4 are organized as follows: \\
Col. 1: Galaxy name, using one or two galaxy identifiers, as in Table 2. 
For Coma, the second
identifier lists the Dressler (1980) catalog number (in the format Dnnn) for 
galaxies that are not included either in the NGC/IC or the CGCG. \\
Cols. 2 and 3: Right Ascension and Declination, in 1950.0 epoch. \\
Col. 4: Morphological type code, in the RC3 scheme (-5 for E, -2 for S0, 0 for
S0a). Morphological types were derived from visual inspection of the blue 
plates of the POSS. Few objects originally classified as S0 had their type
revised to Sa after inspection of the CCD images\\
Col. 5:  Measured apparent I band magnitude. \\
Col. 6:  Radial velocity, in \kms, measured in the CMB reference 
frame. \\
Col. 7:  Angular distance from the cluster center, in degrees. \\
Col. 8:  Measured effective radius, in arcseconds. \\
Col. 9:  Effective radius, in arcseconds, corrected for seeing effects. This is
done adopting the prescriptions of Saglia \etal (1993, see in particular their 
figure 8). \\
Col. 10: Uncertainty on the effective radius, in arcseconds. \\
Col. 11: Corrected metric effective radius, in kiloparsec, obtained 
assuming the galaxy is at the distance indicated by the cluster redshift
and $H_\circ =100$ \kmsm, in the CMB reference frame. \\
Col. 12: Measured effective surface brightness, in I band magnitudes
per square arcsecond. \\
Col. 13: Corrected effective surface brightness, in I band magnitudes per 
square arcsecond. We corrected $\mu_e$ for the smearing effects of seeing, 
adopting the prescriptions of Saglia \etal (1993, see in particular their 
figure 8), for extinction within our Galaxy, using Burstein \& Heiles (1978) 
method as described above, for the cosmological k-correction term (which is 
simply $2.5\log(1+z)$ because of the flat spectrum of early type galaxies in 
the far red), and for the $(1+z)^4$ cosmological dimming. \\
Col. 14: Uncertainty on the effective surface brightness, in magnitudes
per square arcsecond. \\
Col. 15: Stellar velocity dispersion, in \kms. For our new measurements,
the measurement uncertainty, in \kms, is given between brackets: e.g. 177(11) 
is equivalent to 177$\pm$11 \kms. Measurements taken from the literature are
given without uncertainty. See original sources for details. \\
Col. 16: Reference codes for the spectroscopic measurements taken from 
the literature. If an asterisk appears in this column, special comments on the
object are available in section 3.5\\
Whenever multiple spectroscopic measurements were available, we have used 
here only the one with smaller uncertainty. Therefore, if no reference is 
listed in col. 16, the new measurement presented in col. 15 was used for 
that particular galaxy. Otherwise, the measurement in the referenced source, 
and only that one, was used.

Figure 3 shows the spatial distribution (upper panels)
and the radial velocity distribution as a function of angular distance from
the cluster center (lower panels) for the A2634 and Coma FP samples. Symbol
and inset curves follow the same conventions as assumed for Figure 2.

\section{Completeness corrections}

It has been argued that cluster samples for TF and FP work are almost 
bias-free, since all galaxies are roughly at the same distance. While it is
certainly true that the classical Malmquist bias is avoided, a
different and equally important bias affects cluster samples. This is generally
referred-to as the ``cluster population incompleteness bias'' (Teerikorpi 1987,
1990). In a recent work Sandage, Tammann \& Federspiel (1995) have provided a
detailed analysis of this bias and of the corrections it requires before
unbiased distance estimates can be obtained with cluster TF samples.
Their recipes are based on the assumption that galaxy samples are
magnitude-limited. Since that is not the case for our samples, we have derived
incompleteness bias corrections following the spirit of the Sandage \etal 
(1995) treatment, but using Monte Carlo simulations to reproduce as closely as
possible the completeness, measurement uncertainties, and scatter
characteristic of our samples. We refer the reader to the discussion in G96b 
for the details on the bias correction applied to the TF samples, while here we
discuss briefly the bias correction for the FP samples, since the latter is
derived using a significantly different procedure (a more detailed discussion
will be given in Scodeggio 1997).

It is customary to express the completeness of a sample relatively to the 
expected luminosity distribution of a galaxy population, mainly because
magnitudes are relatively easy to measure, and sample selection most often
is based on them. We therefore define completeness as the ratio of the number 
of galaxies of a given magnitude included in a sample with respect to the 
total number of galaxies of the same magnitude predicted by the appropriate 
luminosity function.

The effect of sample incompleteness on the TF relation is relatively
straightforward, because magnitude is one of the two parameters used to
build the relation. At any given velocity width, fainter galaxies are
preferentially missing from an incomplete sample, and this affects the 
determination of the slope, the zero-point, and the dispersion of the 
TF relation. In the FP case, magnitudes enter only indirectly into the FP 
diagram, via the existing relations between luminosity and effective radius 
(Fish 1964), velocity dispersion (Faber \& Jackson 1976), and effective 
surface brightness (Binggeli, Sandage \& Tarenghi 1984). 
These $L - R_e$, $L - \sigma$, and $L - \mu_e$ relations all show rather 
large scatter. Therefore, at any given velocity dispersion or effective
surface brightness, the systematic lack of fainter galaxies is only marginally 
reflected into a lack of small effective radius objects, and it is not 
possible to translate the completeness of a sample 
directly in terms of an effective radius, velocity dispersion or effective 
surface brightness completeness. Our simulations are therefore designed to
reproduce as closely as possible the observed relations between $R_e$ and 
$\sigma$ and the galaxy luminosity, and the relation between $R_e$ and 
$\mu_e$ (Kormendy 1977). Then we select galaxies according to their magnitude,
reproducing the observed luminosity incompleteness, and measure the 
indirect effect of this incompleteness on the FP parameters.
In practice, we first divide both a complete sample and an incomplete one into 
bins according to the value of the combination of $\sigma$ and $\mu_e$ that 
provides an edge-on view of the fundamental plane.
Then we measure the mean value of $R_e$ for each bin, and compare these mean
values for the complete and incomplete samples. Because of the lack of low
luminosity objects, the mean $R_e$ measured in the incomplete sample is
systematically larger than the one measured for the complete sample, except 
in the two to three bins which include the most luminous galaxies. We finally 
use these differences to correct the observed values of $R_e$, and reproduce 
a relation as close as possible to the unbiased one.

\section{The Relative Distance between Coma and A2634}

\subsection{TF}

Although the results of S95 revealed that the A86 result was unlikely to be 
severely affected by substructure in A2634, the A86 TF sample was quite small 
and sparse. The situation improved with the study of G96a, which included a 
sample of 15 strict members and two additional cluster outliers. Our current 
sample includes 23 {\it bona fide} cluster members, and 4 peripheral members, 
nearly tripling the size of the A86 sample. 
The TF relation for our new data is shown
in Figure 4a. Filled symbols indicate cluster members, and unfilled ones
indicate peripheral members. The solid line is the TF template derived by G96b
combining a sample of 555 galaxies in 24 clusters:
\begin{equation}\label{TF_temp}
 M_I = -21.00 (\pm 0.02) - 7.76 (\pm 0.13) \log (W - 2.5)
\end{equation}
The total uncertainty in the TF template zero point is 0.05 mag. 
This includes statistical uncertainties associated with the fit (0.02 mag),
uncertainties associated with the incompleteness bias corrections (0.025 mag),
which are related to the choice of parameters that describe the luminosity
function of spiral galaxies, and the estimated accuracy to within which the
average peculiar velocity of the cluster set approaches zero in the CMB 
reference frame (0.04 mag). An uncertainty of 0.05 mag translates to about 210 
\kms\ at the distance of A2634, and 170 \kms\ at the distance of Coma.
The mean scatter about the TF template relation is $\simeq$0.35 mag., although
this figure varies with the mean velocity width (and luminosity) of galaxies:
it reduces to $\simeq$0.25 mag. at the high $\log W$ end, and expands to
$\simeq$0.45 mag. at the low $\log W$ end of the typical range of the TF
diagram.

The mean offset from the template, for the A2634 sample shown in Fig. 4a, is
--0.04$\pm$0.06 mag., corresponding to a peculiar velocity of +165$\pm$250 
\kms\ in the CMB reference frame. For comparison, the peculiar velocity
estimate obtained by G96b using only 15 strict cluster members is +61$\pm$378
\kms.

For later comparison with the FP results, we also derive a Coma-A2634 
relative TF distance. Figure 4b shows the TF relation for the Coma sample 
of G96a. The convention for the symbols and the solid line are the same as 
in Fig. 4a. The mean offset from the template for Coma is --0.06$\pm$0.06 mag.,
corresponding to a peculiar velocity of +194$\pm$196 \kms. 
The relative peculiar motion of A2634 with respect to Coma is therefore 
-29$\pm$318 \kms, and the ratio of their distances is $1.25\pm0.03$.
We remark also that the differential bias correction between the two clusters 
is only 0.002 magnitudes, and therefore the amplitude of the Coma--A2634 
relative motion in the comoving frame is practically independent of the 
amplitude of the two separate bias corrections.
In the comoving reference frame, it thus appear that the two clusters are at
rest with respect to each other, as nearly determined as the error budget
allows. The peculiar velocities of each of the two clusters also appear to 
be of statistically negligible amplitude.

\subsection{FP}

The \dns\ and FP relations have been obtained in the past using photometric
measurements in many different bands, from B (Faber \etal 1989) to V (L91a, 
L91b), $r_G$ (Djorgovski \& Davis 1987), Gunn r (J{\o}rgensen, 
Franx \& Kj{\ae}rgaard 1996), and Kron-Cousins R (Colless \etal 1993).
In the study of early-type galaxies there are no compelling reasons for 
favoring one optical band over the 
others. Our adoption of I band for photometry was mainly dictated by the 
practical notion that the cluster images contain several galaxies per frame, 
both of early-- and late--type, and thus can be useful for TF application, 
where the I band is favored for compatibility with G96a. 

The FP is a flat surface in the 3-dimensional space of the parameters 
$\log R_e$, $\log \sigma$ , and $\mu_e$. We use the data set of 109 Coma 
galaxies listed in Table 4 in order to obtain a FP template relation, which 
we define as the plane obtained by averaging the coefficients of the 3 
possible fits that can be performed using one of the 3 parameters as the dependent 
variable and the remaining two as the independent ones. 
The resulting FP template, obtained by assuming that Coma is at rest in 
the CMB reference frame and H$_\circ = 100$ \kmsm, is given by
\begin{equation}\label{FP_temp}
\log R_e = 1.25(\pm 0.021)\log \sigma + 0.32(\pm 0.012)~\mu_e - 8.38(\pm 0.008)
\end{equation}
(or $ \log R_e = 1.25 \log \sigma - 0.79 \log I_e - 8.38 $). 
No statistically significant difference is observed when only E or only S0
galaxies are considered. The dispersion around this mean relation, measured 
as the r.m.s. scatter in the residuals of $\log R_e$, is 0.086. 
This is in remarkable agreement with the result of J{\o}rgensen, Franx \& 
Kj{\ae}rgaard (1996), who found a scatter in $\log R_e$ of 0.084. 
The dispersion is equivalent to a scatter of 0.43 
mag in the distance modulus (or 20\% in distance) of a single galaxy.
The statistical uncertainty on the FP zero point is of 0.008, equivalent 
to a distance (or peculiar velocity) uncertainty for the Coma cluster of 133 
\kms.
The incompleteness bias correction for this sample was derived fitting a
Fermi-Dirac distribution to the completeness histogram. The 50\% completeness
level is at $M_I = -20.20$, while the mean I band magnitude of a complete 
sample is $\simeq-20.65$. The mean bias correction to $\log R_e$ derived from 
the Monte Carlo simulations is 0.018 (equivalent to a correction of 0.090 
mag for the distance modulus).

Figure 5 shows an edge-on view of the FP for the Coma sample. The solid line
is the projection of the plane (\ref{FP_temp}). To make the figure more 
readable we plot separately the data points and the error-bars associated with 
those points.
As already pointed out by Djorgovski \& Davis (1987), and J{\o}rgensen, Franx
\& Kj{\ae}rgaard (1996), there might be a small curvature in the plane.
Galaxies with the smallest and largest effective radii appear to have 
systematically positive residuals in $\log R_e$, but the number of objects
involved is too small to make this effect statistically significant.

Figure 6 shows the edge-on view of the FP for the A2634 sample. The 
solid line is the FP template derived for Coma. Error-bars 
are plotted separately in panel (b).  Two galaxies in this sample deviate 
significantly ($> 4 \sigma$) 
from the main relation. One (330747) is a small companion to a much brighter 
galaxy (UGC 12727), so that the iterative subtraction procedure used to obtain
its photometric parameters yields uncertain results.
The second galaxy (331345) has a very flat photometric profile, with abnormally
faint central surface brightness (its effective surface brightness is almost a 
magnitude fainter than the next fainter effective surface brightness in our 
sample). We have excluded these two galaxies from the A2634 sample used to
derive the cluster peculiar motion.

The relative agreement between data-points and template line seen in 
Figure 6 indicates that the difference in peculiar velocity between Coma
and A2634 is small.
The mean offset in $\log R_e$ between A2634 and the Coma template 
(\ref{FP_temp}) is -0.01$\pm$0.013, which corresponds to a peculiar
velocity of --204$\pm$266 \kms\ for A2634, with respect to Coma in the
comoving reference frame. The ratio of the distances of the two clusters,
according to this FP analysis, is $1.27\pm0.03$ This result is in good 
agreement with the estimates obtained using the completely independent
spiral samples and the TF relation, and in disagreement with 
the result of L91a. 
The A2634 sample is characterized by a 50\% completeness level at $M_I = 
-20.70$, and by a bias correction of 0.021 dex. The differential bias 
correction between Coma and A2634 is therefore only 0.003 dex (equivalent to
0.015 magnitudes on the distance modulus), and does not play a significant
role in the determination of the relative motion between the two clusters.

In order to identify the source of the discrepancy with the results of L91a, 
we have re-determined the A2634 peculiar velocity using two subsets of our 
sample. 

The first subset is composed of the 18 galaxies observed by L91a. 
For this computation we used our own I band photometric measurements, and the 
original velocity dispersion measurements of L91a. The mean residual in 
$\log R_e$ for this sample is -0.070$\pm$0.028, corresponding to a 
peculiar velocity of --1324$\pm$573 \kms. This is significantly bigger than 
the value we obtain using the whole sample, but also  smaller than the value 
obtained by L91a using the same velocity dispersions, and their own 
photometric measurements.
As discussed in \S~3.3, from a direct comparison of velocity dispersions the 
L91a measurements appear to be systematically larger, by $\simeq$14\%, than 
our measurements, but this difference alone is not enough to explain
the discrepancy between our distance estimate for A2634 and the L91a one.  
Unfortunately L91a published only V band isophotal diameters for their 
galaxies, and a comparison with our I band 
photometry for those same galaxies is very difficult, without accurate V-I 
color information. Therefore we hesitate to carry the analysis of the L91a 
sample any further. 

We repeat the exercise with a second A2634 subset, composed of the 42 galaxies 
for which we have obtained both photometric and spectroscopic measurements 
(but two galaxies are removed from the sample, as discussed above). 
The mean residual in $\log R_e$ for this sample is 0.000$\pm$0.014, which 
corresponds to a relative peculiar velocity of $0\pm$287 \kms. 
The result is again in good agreement with the TF determination.

The small uncertainties in the TF and FP determinations we have just
discussed, and the good agreement between the two independent results, 
indicate that the peculiar velocity of A2634 is small. 
We conclude that the L91a result is spurious and find no reason to doubt
that the FP and TF relations have universal applicability. 
Similar conclusions have been recently obtained by Lucey and coworkers, 
using new velocity dispersion measurements in A2634 and A2199 (Guzman 1996). 
These new measurements confirm that the measurements reported by L91a were
systematically overestimated, and therefore that the peculiar velocity of
A2634 is negligible. 

\section{Conclusions}

New TF and FP measurements for the cluster A2634 and Coma give compatible
results for the relative distance and peculiar velocity of the two clusters. 
Contrary to the findings of L91a, the peculiar velocity of A2634 with 
respect to the CMB reference frame is unlikely to exceed a few $10^2$
\kms, and extremely unlikely to exceed $10^3$ \kms.

Both TF and FP measurements suggest that the ratio of the distances to
A2634 and to Coma is $\sim 1.26\pm0.03$, which is not too dissimilar
from the ratio of systemic velocities in the CMB reference frame of 1.24.
Our determinations are more accurate and reliable than those of previous
work, thanks to (a) more accurate criteria for the assignment of cluster
membership to individual galaxies and (b) significantly expanded samples.

We have also obtained a new FP template, using for the first time I band 
photometry. The parameters of this template are in very good agreement with
recent determinations of the FP obtained at shorter wavelengths (Gunn r), 
confirming that there is little dependence of the
FP relation on the passband used for the photometric measurements.
The uncertainty with which the FP can provide peculiar velocity 
estimates for single galaxies is $\simeq$0.43 mag in the distance
modulus, or 20\% of the distance. This uncertainty is slightly larger than 
the typical uncertainty that characterizes TF estimates, the latter being
$\simeq$0.35 magnitudes. This disadvantage is however partly
compensated by the fact that the sample incompleteness bias has a less severe
effect on FP cluster distance estimates than it has on TF cluster distance 
estimates, and cluster membership is more readily established for early--type
objects. 

The original motivation for this study was provided by the desire to 
investigate the universality of the TF and FP relations, the 
discrepancy in peculiar velocity estimates reported by L91a making A2634 
an extreme case study. Our results restore a measure of trust in the 
reliability of those relations as cosmological tools. 

\acknowledgments
We would like to thank Juan Carrasco for the precious help he provided during
the observing runs at Palomar, and Bill Schoening for his assistance with the
0.9m telescope at KPNO. We are grateful to the KPNO TAC for the generous
allocation of observing time.
This research was partially supported by NSF grants AST94--20505 to RG, and
AST90--23450 and AST92--18038 to MPH.

\appendix
\section{Comments on individual objects: A2634}

\noindent 330564: marginal cluster membership assignment. \\
331234: probably a member of an interacting system. \\
330636: extremely irregular system, with several clumps of light spread in 
irregular fashion.  Inclination unreliable. \\
330663: spectrum with poor s/n ratio; very uncertain width. \\
U12721: the discrepancy in flux among multiple 
sources reported by G96a is resolved by these observations. \\
U12755: the magnitude reported by G96a is 12.43; new observations 
yield 12.46; we adopt the average value.

\noindent U12716: $\sigma$ measured also by L91a (327 \kms),
Tonry 1984 (348 \kms), Malumuth \& Kirshner 1981 (386 \kms), Tonry \&
Davis 1981 (365 \kms), and  Faber \etal 1989 (298 \kms). \\
331456: $\sigma$ measured also by L91a (218 \kms), and by Tonry 1984 
(250 \kms). \\
330649: $\sigma$ measured also by L91a (285 \kms). \\
330658: $\sigma$ measured also by L91a (201 \kms). \\
330668: $\sigma$ measured also by L91a (207 \kms). \\
330678: $\sigma$ measured also by L91a (272 \kms). \\
331456, 330648, 330651, 330658, 330660, 330665, 330668, 330678 are within the 
halo of the cD UGC 12716; their $\mu_e$ has high uncertainty. \\
330706: bright nearby star subtracted. \\
330600: faint outer disk visible. \\
330747: bright nearby companion (UGC 12727) subtracted. Surface brightness
determination unreliable. \\
U12744: bright nearby star subtracted.

\section{Comments on individual objects: Coma}

\noindent U8049: $\sigma$ also measured by L91b (212 \kms), 
and Faber \etal 1989 (208 \kms). \\
 U8065: $\sigma$ also measured by L91b (279 \kms). \\
 U8070: $\sigma$ also measured by Davies \& Illingworth 1983 (255 \kms), Faber
\etal 1989 (259 \kms), and L91b (298 \kms). \\
 U8103: $\sigma$ also measured by L91b (279 \kms), Faber \etal 1989 (245 
\kms), Malumuth \& Kirshner 1981 (304 \kms). Ours spectrum has poor s/n ratio, 
and therefore we use Faber \etal measurement. \\
 U8110: $\sigma$ also measured by Oegerle \& Hoessel 1991 (381 \kms), Tonry 
1985 (404 \kms), Tonry \& Davis 1981 (412 \kms), Davies \& Illingworth 1983 
(342 \kms), Faber \etal 1989 (381 \kms), Faber \& Jackson 1976 (400 \kms),
and L91b (414 \kms). \\
 U8175: $\sigma$ also measured by L91b (275 \kms). \\
221266: $\sigma$ also measured by Davies \etal 1987 (150 \kms). \\
221354: $\sigma$ also measured by L91b (147 \kms), and Faber et 
al. 1989 (164 \kms). \\
221410: $\sigma$ also measured by Dressler 1987 (133 \kms). \\
221216: spectrum with poor s/n ratio. Also, the galaxy is in a close
pair. \\
221290, 221291, 221293, 221298, 221303, 221304, 221317, 221323, 221329,
221331, 221334 are within the halo of the D galaxy UGC 8103, and
221354, 221362, 221377, 221380, 221382, 221392 are within the halo of the D 
galaxy UGC 8110; their $\mu_e$ has high uncertainty. \\
 U8133: faint outer disk. \\
 U8072/U8073: close pair; iterative subtraction required. \\
 U8100: bright nearby star subtracted.

\newpage

\figcaption[a34_fig1.ps]{Comparison of our velocity dispersion measurements 
with literature data. The difference (our -- literature) is plotted vs. 
our measurements.
Filled symbols: Coma cluster galaxies, with measurements from L91b, Davies 
\etal (1987), and Faber \etal (1987). Empty symbols: A2634 galaxies, with
measurements from L91a. The lines connecting pairs of data-points identify
3 galaxies for which multiple comparisons are available.}

\figcaption[a34_fig2.ps]{Spatial distribution (upper panels) and cz$_{CMB}$ vs.
angular separation from the cluster center (lower panels) for the A2634 and
Coma TF samples. Filled circles indicate galaxies that are considered
cluster members; empty circles indicate peripheral cluster member; asterisks
indicate members of foreground or background groups (see discussion in the 
text). Small symbols indicate all galaxies with measured redshift located in 
the area. The two dashed circles (upper panels), and the two
dashed vertical lines (lower panels) refer to distances of 1 and 2 $R_A$ 
from the cluster centers. The horizontal lines in the lower panels 
mark the cluster systemic velocity in the CMB reference frame. The curved solid
lines are the redshift-space caustic lines estimated for $\Omega_\circ = 0.3$.}

\figcaption[a34_fig3.ps]{FP samples, plotted in the same mode and convention 
as described for figure 1.}

\figcaption[a34_fig4.ps]{(a) TF diagram for A2634. (b) TF 
diagram for Coma. In both figures filled and empty symbols indicate cluster
members and peripheral members, respectively. 
The solid line is the G96b TF template (equation 1).}

\figcaption[a34_fig5.ps]{Edge-on view of the FP for Coma. 
The upper panel shows the data points, and the projection of the best fitting 
plane (equation 2). The lower panel shows the error bars
associated with the upper panel data points.}

\figcaption[a34_fig6.ps]{Edge-on view of the FP for A2634. The inset straight 
line is the FP template relation computed for the Coma sample.
Filled and empty symbols indicate cluster members and peripheral members, 
respectively, and asterisks indicate the possible members of foreground and
background groups (see discussion in the text).}

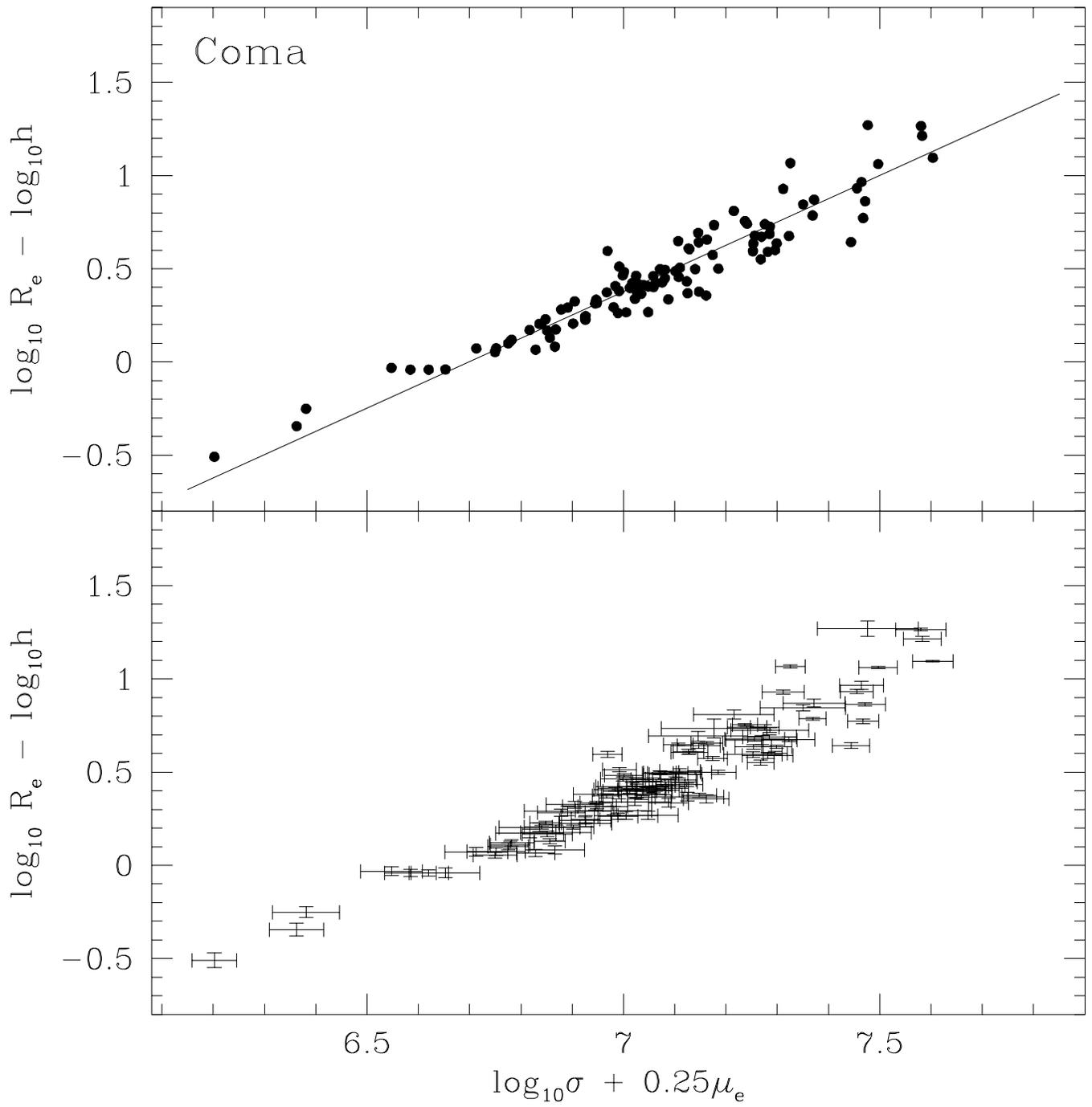
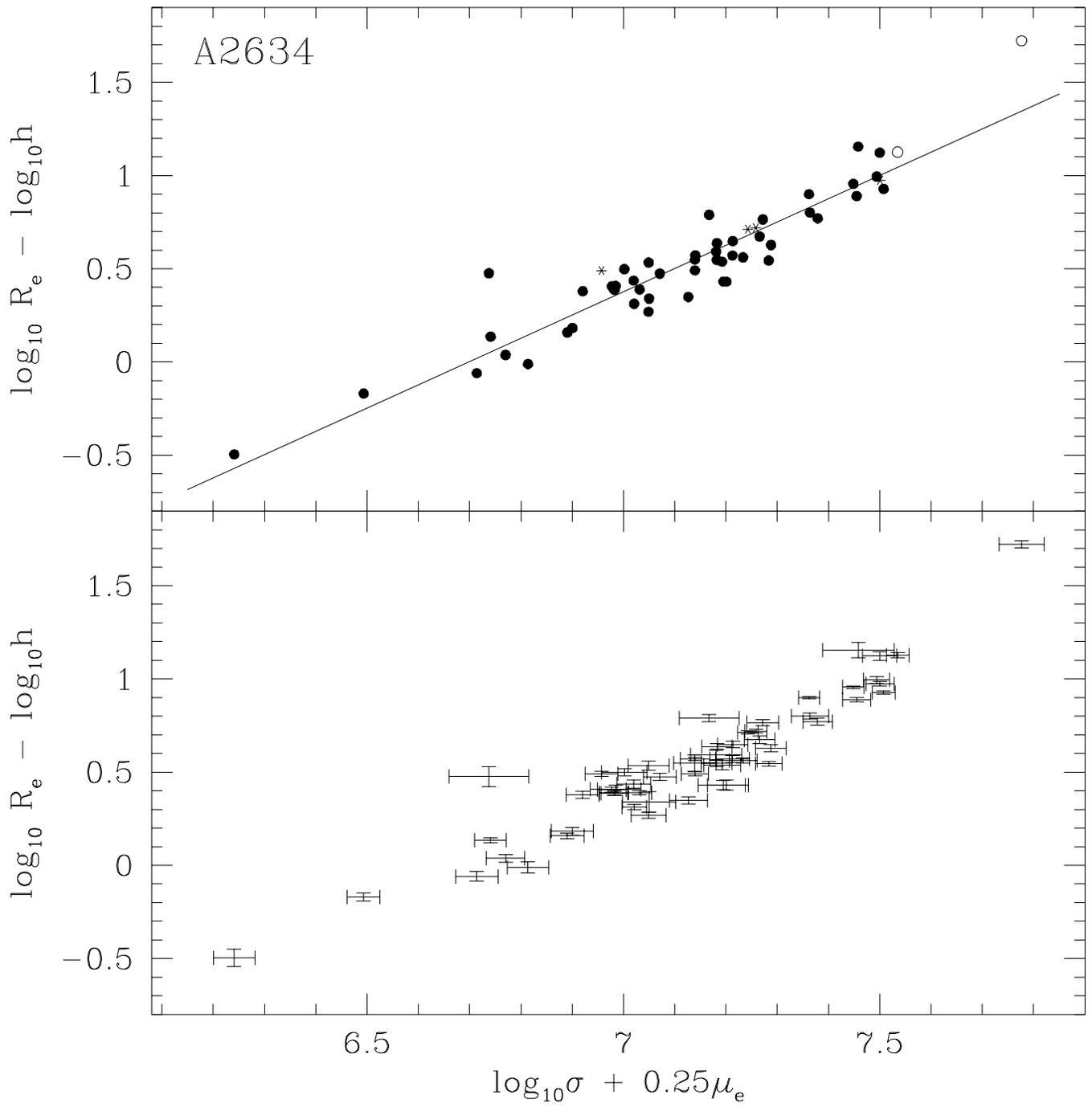
\end{document}